\documentclass[aps,prb,twocolumn,a4paper,10pt]{revtex4-2}
\usepackage{amssymb}
\usepackage{amsmath}
\usepackage{graphicx}
\usepackage{color}
\usepackage{subfigure}
\usepackage{float}

\begin{document}

\title{Temporal and spatial attenuation of inertial spin waves driven by spin-transfer torques}
\date{\today}

\author{Peng-Bin He$^1$\footnote[0$^{\dag}$]{$\dag$ hepengbin@hnu.edu.cn}, and Mikhail Cherkasskii$^2$\footnote[0$^{\star}$]{$\star$ macherkasskii@hotmail.com}}

\affiliation{$^1$School of Physics and Electronics, Hunan University, Changsha 410082, China \\
$^2$Institute for Theoretical Solid State Physics, RWTH Aachen University, DE-52074 Aachen, Germany}

\begin{abstract}
Magnetic damping induces the temporal and spatial decay of spin waves, characterized by the damping factor and attenuating length, both of which can be measured to determine various magnetic and spin-transport parameters. By investigating the dispersion and dissipation of inertial spin waves driven by spin-transfer torques, we find that magnetic inertia modifies the dependence of the damping factor and attenuating length on the electric current and spin wave frequency. This provides a valuable method for probing the inertial relaxation time.
\end{abstract}

\maketitle

\section{introduction} \label{int}

The recently discovered magnetic inertia can promote conventional ferromagnets to materials capable of operating in the terahertz range \cite{NeerajK}. Experimentally detected \cite{NeerajK,LiY,UnikandanunniV,DeA} magnetic inertia manifests as nutation superimposed on the precessional motion of magnetic moments. Nutation can be viewed as ultrafast oscillations around the precessional trajectory \cite{MondalR_JMMM,OliveE_APL,BottcherD,OliveE_JAP,CherkasskiiM_PRB102,TitovSV_PRB103,CiorneiMC,WegroweJE}. These oscillations arise due to the non-vanishing inertial relaxation time, which prevents magnetic moments from
instantaneously following angular momenta \cite{WegroweJE,WinterL}. In addition to being the origin of the (sub-)terahertz mode, nutation causes a redshift in the precessional motion in ferromagnets \cite{OliveE_APL,OliveE_JAP,CherkasskiiM_PRB106,TitovSV_JAP}. Unfortunately, this shift is difficult to detect experimentally, as it is challenging to distinguish the influence of inertia from other factors, such as anisotropy.

Notably, inertial spin dynamics has also been observed in antiferromagnets a decade ago \cite{KimelAV}, though it originates from inter-lattice interactions. The mathematical isomorphism between antiferromagnetic and pure
magnetic inertia was established later \cite{RodriguezR}. Beyond that, magnetic inertia was predicted in antiferromagnets \cite{MondalR_PRB103,MondalR_PRB104} and in complex spin structures \cite{CherkasskiiM_PRB109}, further  enriching their underlying physics. Magnetic inertia doubles the number of modes in antiferromagnets, with nutational  modes predicted to exhibit high intensity and sharp lineshapes \cite{MondalR_PRB103}. However, to the best of our knowledge, there has been no experimental detection of magnetic inertia in antiferromagnets.

Despite the significant impact of magnetic inertia on magnetization dynamics, determining its key parameter -- the inertial relaxation time -- remains a challenging problem for both experimentalists and theoreticians. This
challenge can be addressed experimentally using phase-resolved magneto-optical Kerr effect techniques \cite{NeerajK,UnikandanunniV} or high-frequency ferromagnetic resonance \cite{LiY}. Theoretically, it can be calculated from first principles, involving s-d-like interactions \cite{BhattacharjeeS} or through the torque-torque correlation model \cite{ThonigD}. The lack of simple experimental and theoretical methods limits the understanding of magnetic
inertia in both ferromagnetic and antiferromagnetic materials.

Spin waves and their decay are subjects of intensive investigation in the field of magnonics \cite{FlebusB}. The decay can be controlled and suppressed by nonadiabatic spin-transfer torque (STT) \cite{SeoSM,MoonJH,KimDH,ZhouZW,ChauleauJY}, Slonczewski's STT \cite{XingXJ,WooS} and a combination of both \cite{YanZM} due to their anti-damping properties. Notably, spin-wave decay can be utilized to estimate the damping factor \cite{ManagoT}, detect the nonadiabaticity of STTs \cite{SeoSM,ChauleauJY}, characterize the Dzyaloshinskii-Moriya interaction \cite{MoonJH}, measure the angular momentum compensation point of ferrimagnets \cite{KimDH}, and control spin-wave polarization \cite{ZhouZW}.

It has been predicted that magnetic inertia affects the dispersion of precessional spin waves and can give rise to nutational spin waves, with their frequency being inversely proportional to the inertial relaxation time \cite{MakhfudzI_APL,CherkasskiiM_PRB103,LomonosovAM,TitovSV_PRB105,MondalR_PRB106,CherkasskiiM_PRB109}. Recent studies have demonstrated that Slonczewski's STT generates nutational auto-oscillations in inertial ferromagnets \cite{RodriguezR} and antiferromagnets \cite{HePB_PRB108,HePB_PRB110}. However, the influence of STTs on spin waves has been scarcely addressed. Additionally, since spin-wave decay and STTs have been successfully used to reveal the properties of precessional spin waves, it is proposed that decay could also serve as a method to investigate nutational spin waves and, consequently, magnetic inertia.

In resonance experiments, temporal decay is typically the focus, characterized by the damping factor \cite{LomonosovAM,ZhouZW} and described by an exponential time decay. In contrast, spin-wave transport experiments are more concerned with the spatial attenuating length \cite{SeoSM,MoonJH,KimDH,ChauleauJY,XingXJ,WooS,YanZM,WangJ}, which features an exponential space decay. By leveraging the dependence of the damping factor and attenuating length on magnetic inertia in spin waves driven by STTs, our study offers a promising approach to estimate the inertial relaxation time, which signifies the characteristic time scale of inertial dynamics.

The paper is organized as follows. After the introduction, in Sec. \ref{int} the model and method are presented in Sec. \ref{mod}. The eigenmodes of initial spin waves are briefly discussed in Sec. \ref{eig}. The temporal and spatial aspects of spin-wave decay are examined in detail in Secs. \ref{tem} and \ref{spa}. Finally, Sec. \ref{con} summarizes the conclusions.

\section{model and method} \label{mod}

\begin{figure}[b]
\includegraphics[scale=0.3,angle=0,trim=0cm 0.5cm 0cm 0.5cm]{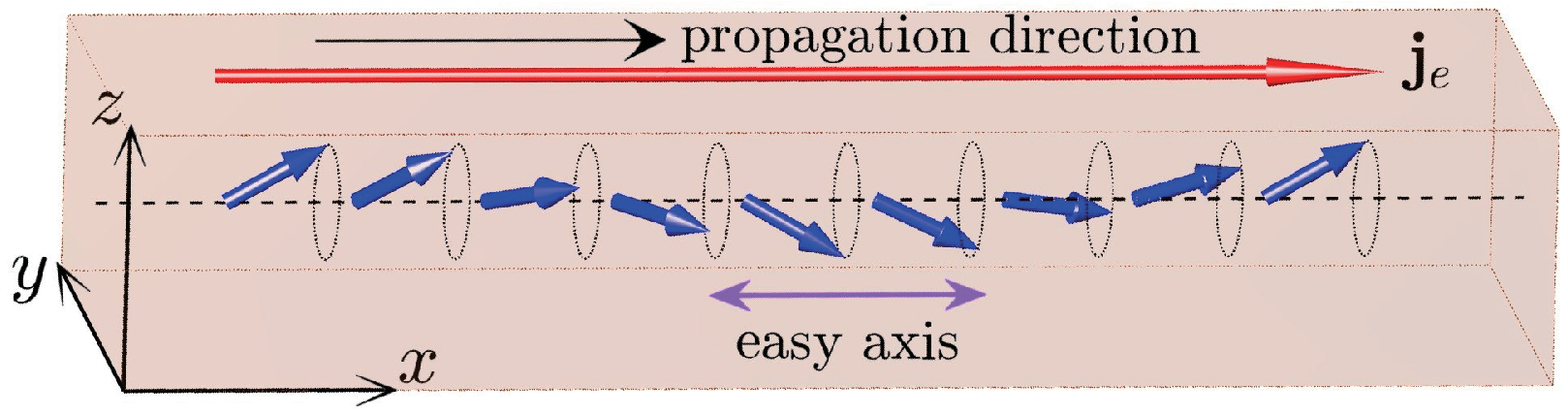}
\caption{(color online). Schematic diagrams of model system.}
\label{model}
\end{figure}

We consider ferromagnetic stripes driven by the spin-transfer torques (STTs) \cite{ZhangS}, as shown in Fig. \ref{model}. We extend the inertial Landau-Lifshitz-Gilbert (ILLG) equation \cite{CiorneiMC,WegroweJE} by incorporating STT terms, i.e., \begin{eqnarray}
\frac{\partial \mathbf{m}}{\partial t} &=& - \gamma \mathbf{m} \times \mathbf{H}_{eff} + \alpha \mathbf{m} \times \frac{\partial \mathbf{m}}{\partial t} + \eta \mathbf{m} \times \frac{\partial^2 \mathbf{m}}{\partial t^2} + \mathbf{T}, \notag \\ \label{iLLG}
\end{eqnarray}
where $\mathbf{m}$ is the unit vector of magnetization. The gyromagnetic ratio reads $\gamma = g \mu_0 \mu_B/\hbar$, with $g$ being the Land\'{e} $g$-factor, $\mu_0$ the vacuum susceptibility, $\mu_B$ the Bohr magneton, and $\hbar$ the reduced Plank constant. In Eq. (\ref{iLLG}), $\alpha$ is the Gilbert damping constant, and $\eta$ is the inertial relaxation time ranging from fs to ps, as predicted in the \textit{ab initio} calculation \cite{ThonigD} and measured in the nutation experiments \cite{LiY,NeerajK,UnikandanunniV}.

The effective field is expressed as the functional derivative of magnetic energy, namely, $\mathbf{H}_{eff} = - [1/(\mu_0 M_s)] \delta E/\delta \mathbf{m}$ with $M_s$ being the saturation magnetization. The magnetic energy, consisting of contributions from the exchange interaction, as well as the anisotropy and the demagnetization fields, is given by
\begin{eqnarray}
E &=& \int d r^3 \Big \{ A \left( \nabla \mathbf{m} \right)^2 - K \left( \mathbf{m} \cdot \mathbf{e}_x \right)^2 \notag \\ && + \frac{1}{2} \mu_0 M_s^2 \left[ N_y \left( \mathbf{m} \cdot \mathbf{e}_y \right)^2 + N_z \left( \mathbf{m} \cdot \mathbf{e}_z \right)^2 \right] \Big \}, \label{energy}
\end{eqnarray}
with $A$ being the exchange stiffness, $K$ the anisotropy constant, and $N_{y,z}$ the demagnetization factor.

The adiabatic and nonadiabatic STTs read
\begin{equation}
\mathbf{T} = \left( \mathbf{u} \cdot \nabla \right) \mathbf{m} - \beta \mathbf{m} \times \left[ \left( \mathbf{u} \cdot \nabla \right) \mathbf{m} \right], \label{STT}
\end{equation}
where $\beta$ denotes the nonadiabaticity of STTs. The spin transport is integrated into a magnetization drift velocity,
\begin{equation}
\mathbf{u} = P \frac{\mu_B}{e M_s} \mathbf{j}_e,
\end{equation}
with $P$ being the spin polarization and $\mathbf{j}_e$ the current density.

We focus on a stripe with a cross-section significantly smaller than its length. The easy magnetic axis lies along the stripe (see Fig. \ref{model}), therefore the anisotropic field ensures a uniform magnetization distribution at equilibrium. The current and the small-amplitude spin wave propagate along the stripe as well. Hence, the substitution of  Eqs. (\ref{energy},\ref{STT}) into Eq. (\ref{iLLG}) yields the explicit form of the ILLG equation
\begin{eqnarray}
\frac{\partial \mathbf{m}}{\partial t} &=& - \gamma J \mathbf{m} \times \frac{\partial^2 \mathbf{m}}{\partial x^2} - \gamma H_k m_x \left( \mathbf{m} \times \mathbf{e}_x \right) \notag \\ && + \gamma N_y M_s m_y \left( \mathbf{m} \times \mathbf{e}_y \right) + \gamma N_z M_s m_z \left( \mathbf{m} \times \mathbf{e}_z \right) \notag \\ && + \alpha \mathbf{m} \times \frac{\partial \mathbf{m}}{\partial t} + \eta \mathbf{m} \times \frac{\partial^2 \mathbf{m}}{\partial t^2} \notag \\ && + u \frac{\partial \mathbf{m}}{\partial x} - \beta u \mathbf{m} \times \frac{\partial \mathbf{m}}{\partial x}, \label{explicit_iLLG}
\end{eqnarray}
where $J = 2 A/(\mu_0 M_s)$, and $H_k = 2 K/(\mu_0 M_s)$.

We linearize Eq. (\ref{explicit_iLLG}) by expanding $\mathbf{m}(x,t)$ around the equilibrium state with the plane-wave ansatz
\begin{equation}
\mathbf{m} = \pm \mathbf{e}_x + \left( A_y \mathbf{e}_y + A_z \mathbf{e}_z \right) e^{i \left( \tilde{k} x - \tilde{\omega} t \right)}, \label{ansatz}
\end{equation}
where $A_{x,y}$ are the amplitudes. Eq. (\ref{explicit_iLLG}) can be analyzed from different perspectives depending on the characteristics of $\tilde{\omega}$ and $\tilde{k}$. If the frequency and wave number are real, the properties of eigen spin waves in a conservative system can be explored. However, it is insightful to consider temporally and spatially damped spin waves separately by assuming either a complex frequency with a real wave number or vice versa. In this approach, the real (imaginary) part of $\tilde{\omega}$ and $\tilde{k}$ correspond to the dispersion (dissipation) of spin waves. From the experimental point of view, the imaginary part of $\tilde{\omega}$ is associated with the linewidth in ferromagnetic resonance measurement. Conversely, in spin-wave transport experiments, the attenuating length can be determined as the inverse of the imaginary part of $\tilde{k}$ (see Refs. [\onlinecite{SeoSM,MoonJH,KimDH,ChauleauJY,XingXJ,WooS,YanZM,WangJ})]. In the present theoretical study, we first consider the eigen spin waves, then we assume the real wave number and complex frequency. In the third part, we focus on the complex $\tilde{k}$ and real $\tilde{\omega}$.

Substituting Eq. (\ref{ansatz}) into Eq. (\ref{explicit_iLLG}), and keeping the linear terms on $A_x$ and $A_y$, one has
\begin{equation}
\left( \begin{array}{cc} a_{11} - \tilde{\omega} & a_{12} \\ a_{21} & a_{22} - \tilde{\omega} \end{array} \right) \left( \begin{array}{c} A_y \\ A_z \end{array} \right) = 0. \label{linearized_Eq}
\end{equation}
where $a_{11} = a_{22} = - u \tilde{k}$, $a_{12} = i \eta \tilde{\omega}^2 - \alpha \tilde{\omega} - \beta u \tilde{k} - i \omega_z ( \tilde{k} )$, and $a_{21} = - i \eta \tilde{\omega}^2 + \alpha \tilde{\omega} + \beta u \tilde{k} + i \omega_y ( \tilde{k} )$, with $\omega_{y,z}(x) = \gamma \left( N_{y,z} M_s + H_k + J x^2 \right)$. To ensure a nontrivial solution, the determinant of the coefficient matrix must be zero. Then one can get a secular equation
\begin{eqnarray}
\eta^2 \tilde{\omega}^4 + 2 i \alpha \eta \tilde{\omega}^3 + c_2 \tilde{\omega}^2 + c_1 \tilde{\omega} + c_0 = 0, \label{secular_eq}
\end{eqnarray}
where $c_2 = - ( 1 + \alpha^2 ) - \eta [ \omega_y ( \tilde{k} ) + \omega_z ( \tilde{k} ) -  2 i \beta u \tilde{k} ]$, $c_1 = - 2 ( 1 + \alpha \beta ) u \tilde{k} - i \alpha [ \omega_y ( \tilde{k} ) + \omega_z ( \tilde{k} )  ]$, and $c_0 = \omega_y ( \tilde{k} ) \omega_z ( \tilde{k} ) - ( 1 + \beta^2 ) u^2 \tilde{k}^2 - i \beta u [ \omega_y ( \tilde{k} ) + \omega_z ( \tilde{k} ) ] \tilde{k}$. Based on Eq. (\ref{secular_eq}), the following sections discuss the eigen inertial spin waves, temporally damped spin waves, and spatially attenuating spin waves.

\section{eigen inertial spin waves} \label{eig}

By neglecting damping and STTs ($\alpha = 0$ and $u = 0$), Eq. (\ref{explicit_iLLG}) describes a conservative system with real $\tilde{k}$ and $\tilde{\omega}$. Consequently, solving Eq. (\ref{secular_eq}) yields four different frequencies in the presence of a non-vanishing inertial relaxation time ($\eta \neq 0$). Two of them ($\pm \omega_p$) are associated with the precessional spin waves that exists even without nutation. The other two ($\pm \omega_n$) represent the nutational spin waves with higher frequencies. $\omega_{n,p}$ are written as
\begin{small}
\begin{equation}
\omega_{n,p} = \frac{\sqrt{1 + \eta \left( \omega_y + \omega_z \right) \pm \sqrt{\left[ 1 + \eta \left( \omega_y + \omega_z \right) \right]^2 - 4 \eta^2 \omega_y \omega_z}}}{\sqrt{2} \eta}, \label{eigen_fre}
\end{equation}
\end{small}
where
\begin{equation}
\omega_{y,z} = \gamma \left( N_{y,z} M_s + H_k + J k^2 \right), \label{omega_yz_2}
\end{equation}
with $k$ denoting the magnitude of wave vector. Note that the eigenfrequencies come in two pairs $\pm \omega_{n,p}$. Eq. (\ref{eigen_fre}) gives the positive frequencies. In the limit of $\eta \rightarrow 0$, $\omega_n$ approaches infinity, and $\omega_p$ approaches $\sqrt{\omega_y \omega_z}$, which is the eigenfrequency of non-inertial spin waves. Hence, $\omega_n$ and $\omega_p$ belong to the nutational and precessional spin waves respectively. For small $\eta$, the Taylor series expansion results in the approximate frequencies,
\begin{eqnarray}
\omega_n &=& \frac{1}{\eta} + \frac{\omega_y + \omega_z}{2} + O[\eta], \label{eigen_nut_fre} \\
\omega_p &=& \sqrt{\omega_y \omega_z} \left( 1 - \eta \frac{\omega_y + \omega_z}{2} \right) + O[\eta]^2. \label{eigen_pre_fre}
\end{eqnarray}
The structure of Eqs. (\ref{eigen_nut_fre},\ref{eigen_pre_fre}) is similar to the results presented in Ref. [\onlinecite{LomonosovAM}]. The magnetic inertia gives rise to nutational spin waves, and is responsible for the red shift of the original precessional spin waves.

\section{temporal damping spin waves} \label{tem}

\begin{figure*}[t]
\includegraphics[scale=0.26,angle=0,trim=0cm 3cm 0cm 0cm]{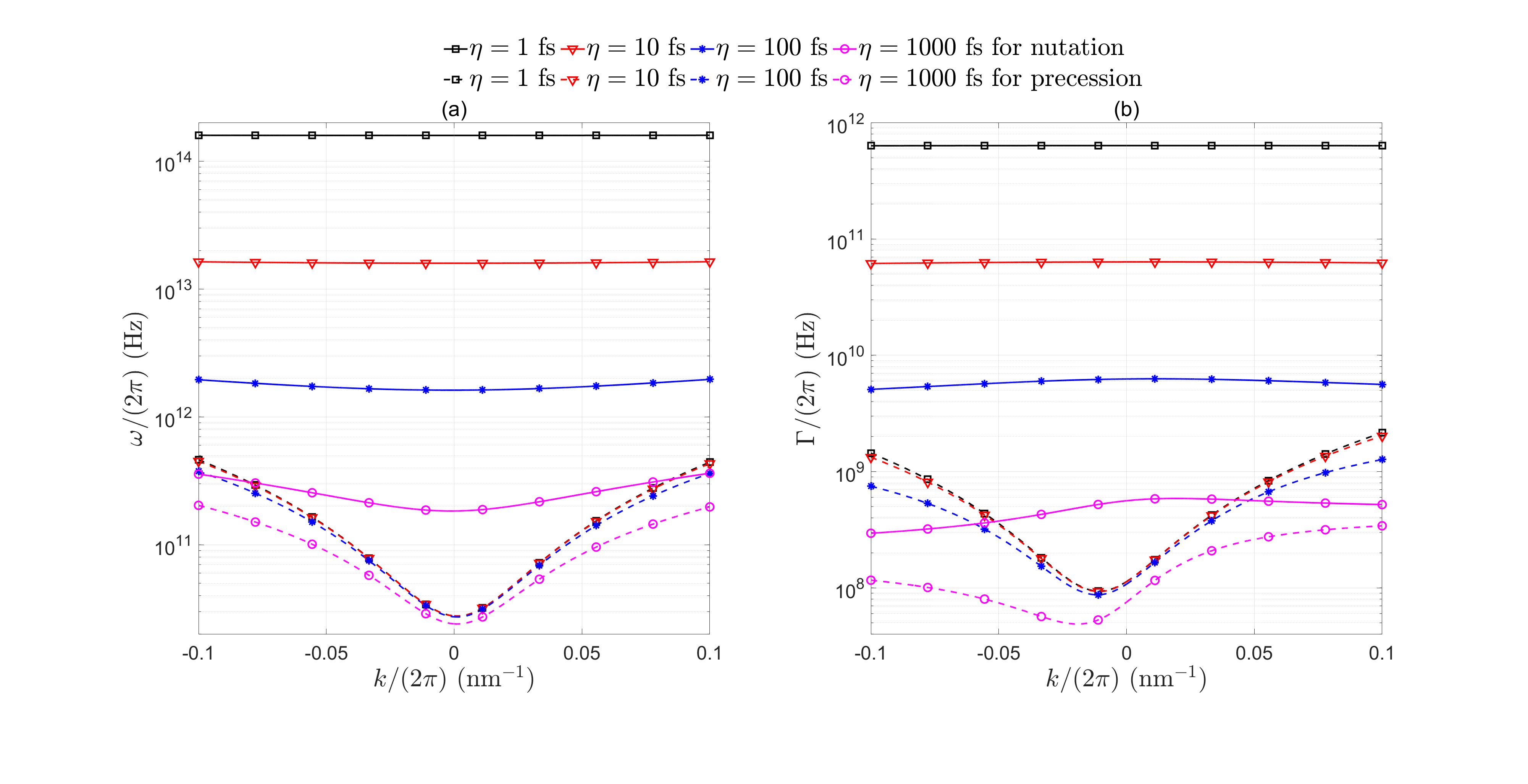}
\caption{(color online). Dispersion (a) and dissipation (b) for nutational and precessional spin waves. The magnetic parameters are adopted from CoFe alloys \cite{LiuX,WeberR}: the exchange stiffness $A = 30.2$ pJ$/$m, the anisotropy constant $K = 0.02$MJ$/$m$^3$, the saturation magnetization $M_s = 1.57$ MA$/$m, and the Gilbert damping constant $\alpha = 0.004$. The demagnetizing factors are chosen as $N_y = 0.6$ and $N_z = 0.4$. The parameters of STTs are $\beta = 0.04$ and $u = 100 $m$/$s.} \label{omega_and_Gamma_vs_k}
\end{figure*}

From the standpoint of the temporal damping of spin waves, it is assumed that $\tilde{k} = k$ is real, but the frequency is complex,
\begin{equation}
\tilde{\omega} = \omega - i \Gamma, \label{complex_omega}
\end{equation}
with $\omega$ being the real part, and $\Gamma$ the damping factor. According to the ansatz [(Eq. (\ref{ansatz})], Eq. (\ref{complex_omega}) yields an exponential time decay $\exp (- \Gamma t)$. By substituting Eq. (\ref{complex_omega}) into the secular equation (\ref{secular_eq}), and separately matching the real and imaginary parts respectively, two higher-order polynomial equations with respect to $\omega$ and $\Gamma$ are obtained,
\begin{small}
\begin{eqnarray}
&& \left[ \frac{\omega_y + \omega_z}{2} - \alpha \Gamma - \eta \left( \omega^2 - \Gamma^2 \right) \right]^2 - \left[ \left( \alpha \omega + \beta u k \right) - 2 \eta \omega \Gamma \right]^2 \notag \\ && + \Gamma^2 - \left( \omega + u k \right)^2 = \left[ \frac{\omega_y - \omega_z}{2} \right]^2, \label{re_eq_tem} \\
&& \left[ \frac{\omega_y + \omega_z}{2} - \alpha \Gamma - \eta \left( \omega^2 - \Gamma^2 \right) \right]  \left[ \left( \alpha \omega + \beta u k \right) - 2 \eta \omega \Gamma \right] \notag \\ && = \Gamma \left( \omega + u k \right). \label{im_eq_tem}
\end{eqnarray}
\end{small}
There is no exact closed-form solution of Eqs. (\ref{re_eq_tem}, \ref{im_eq_tem}). Therefore, we first solve them numerically.

By selecting $\eta$ within the reliable range (Ref. [\onlinecite{MondalR_JMMM}]), Fig. \ref{omega_and_Gamma_vs_k} illustrates the frequency and the damping factor as functions of $k$ for nutational and precessional spin waves. Several key features can be observed in Fig. \ref{omega_and_Gamma_vs_k}. First, in panel (a), the $\omega$-$k$ curves for nutation are flatter compared to those for precession, indicating that the dispersion of precessional spin waves is more pronounced than that of nutational spin waves. This implies that the group velocity of precessional spin waves is higher than that of nutational spin waves.

Second, both the frequency and the damping factor are significantly influenced by magnetic inertia. As $\eta$ increases, both precessional and nutational spin waves experience a redshift, and their damping factors decrease. Additionally, the dispersion of nutational spin waves becomes more pronounced with increasing $\eta$, while the dispersion of precessional spin waves weakens.

Third, STTs bring the nonreciprocity in the propagation of spin waves. In Fig. \ref{omega_and_Gamma_vs_k}, we take moderate values of $u$ and $\beta$, for which the nonreciprocity is barely visible in Fig. \ref{omega_and_Gamma_vs_k}(a). However, the $\Gamma$-$k$ curves [Fig. \ref{omega_and_Gamma_vs_k}(b)] exhibit clear asymmetry under the inversion $k \rightarrow - k$. Note that we observed induced nonreciprocity in contrast to inherit Dzyaloshinskii-Moriya nonreciprocity, which appears due to broken inversion symmetry. On the other hand, flipping the sign of $k$ is equivalent to reversing the current direction $u\rightarrow-u$ [see Eqs. (\ref{re_eq_tem}) and (\ref{im_eq_tem})]. The nonreciprocity can be utilized to determine the inertial relaxation time $\eta$, and we discuss this opportunity further below.

To gain insights into the underlying physics, we temporarily switch our focus on the special case of uniaxial magnetic anisotropy. The right-hand-side term of Eq. (\ref{re_eq_tem}), proportional to $N_y - N_z$, complicates obtaining an exact analytical solution for Eqs. (\ref{re_eq_tem}) and (\ref{im_eq_tem}). For an infinitely long magnetic stripe, the demagnetization factors depend on the aspect ratio of the stripe \cite{AharoniA}. If the aspect ratio is $1$, $N_y = N_z = 1/2$, resulting in $\omega_y = \omega_z = \gamma ( 1/2 M_s + H_k + J k^2 )$, which causes the right-hand-side term to vanish. Consequently, Eqs. (\ref{re_eq_tem}) and (\ref{im_eq_tem}) can be solved analytically.

By taking $\omega_y = \omega_z$ in Eqs. (\ref{re_eq_tem}, \ref{im_eq_tem}), one obtains four dispersion branches under the approximation $1 + \alpha^2 \approx 1$,
\begin{eqnarray}
\omega_p^\pm = \pm \frac{\sqrt{1 + 4 \eta \Omega_\mp + \sqrt{\left( 1 + 4 \eta \Omega_\mp \right)^2 + 4 \alpha_\mp^2}} - \sqrt{2}}{2 \sqrt{2} \eta}, \label{omega_p} \\
\omega_n^\pm = \pm \frac{\sqrt{1 + 4 \eta \Omega_\pm + \sqrt{\left( 1 + 4 \eta \Omega_\pm \right)^2 + 4 \alpha_\pm^2}} + \sqrt{2}}{2 \sqrt{2} \eta}, \label{omega_n}
\end{eqnarray}
where
\begin{eqnarray}
\alpha_\pm &=& \alpha \pm 2 \eta \beta u k, \\
\Omega_\pm &=& \omega_0 + \gamma J \left( k \pm k_0 \right)^2, \label{Omega_pm}
\end{eqnarray}
with $\omega_0 = \gamma (1/2 M_s + H_k) - u^2/(4 \gamma J)$, and $k_0 = u/(2 \gamma J)$. The corresponding damping factors can be written as,
\begin{eqnarray}
\Gamma_p^\pm &=& \frac{\alpha \omega_p^\pm + \beta u k}{2 \eta \omega_p^\pm \pm 1}, \label{Gamma_p} \\
\Gamma_n^\pm &=& \frac{\alpha \omega_n^\pm + \beta u k}{2 \eta \omega_n^\pm \mp 1}. \label{Gamma_n}
\end{eqnarray}
Equations (\ref{omega_p})-(\ref{Omega_pm}) explicitly reveal the influences of STTs on the dispersion. In the absence of nonadiabatic STT, the dispersion curve is shifted along the $k$-axis due to the term $(k \pm k_0)^2$, and the $\omega$-axis due to the $u^2/(4 \gamma J)$ term. In other words, the adiabatic STT alters both the frequency and the wavelength. This behavior is reminiscent of the well-known Doppler effect of spin waves \cite{VlaminckV,ChauleauJY,SugimotoS}. The terms $\alpha \omega_{p,n}^\pm + \beta u k$ in Eqs. (\ref{Gamma_p}, \ref{Gamma_n}) represent the competition between the nonadiabatic STT and the intrinsic damping. This competition can either weaken or enhance spin-wave decay, depending on the relative directions of the current and the spin wave.

For small $\eta$, one can find the approximation for precession and nutation frequencies
\begin{eqnarray}
\omega_p^\pm &=& \pm \Omega_\mp - \alpha \beta u k \mp \eta \left \{ \left( \Omega_\mp \mp \alpha \beta u k \right)^2 \right.  \notag \\ && \left. - \left[ 2 \alpha \left( \Omega_\mp \mp \alpha \beta u k \right) \pm \beta u k \right]^2 \right \} + O[\eta]^2, \label{omega_p_approx} \\
\omega_n^\pm &=& \pm \frac{1}{\eta} \pm \Omega_\pm + \alpha \beta u k + O[\eta] \label{omega_n_approx},
\end{eqnarray}
In the absence of damping and STTs, Eqs. (\ref{omega_p_approx}) and (\ref{omega_n_approx}) are respectively reduced to Eqs. (\ref{eigen_pre_fre}) and (\ref{eigen_nut_fre}) by setting $N_x = N_y = 1/2$. In non-inertial case, $\omega_{n}^{\pm}$ approach infinity, indicating the disappearance of nutational spin waves. Similarly, for precessional spin waves, $\omega_p^\pm \rightarrow \pm \Omega_\mp - \alpha \beta u k$, and $\Gamma_p^\pm \rightarrow \alpha \Omega_\pm \pm \beta u k$. By substituting Eq. (\ref{Omega_pm}) into these expressions and approximating $1 + \alpha \beta$ to $1$, we recover the results for spin waves in uniaxial ferromagnets \cite{ZhouZW}. This confirms the validity of our calculations.  It is worth emphasizing that the dispersion [see Eqs. (\ref{omega_p}, \ref{omega_n})] and the dissipation [see Eqs. (\ref{Gamma_p}, \ref{Gamma_n})] are the same as those of the uniaxial ferromagnet, although they correspond to a special wire geometry here.

\begin{figure*}[t]
\includegraphics[scale=0.26,angle=0,trim=0cm 4cm 0cm 0cm]{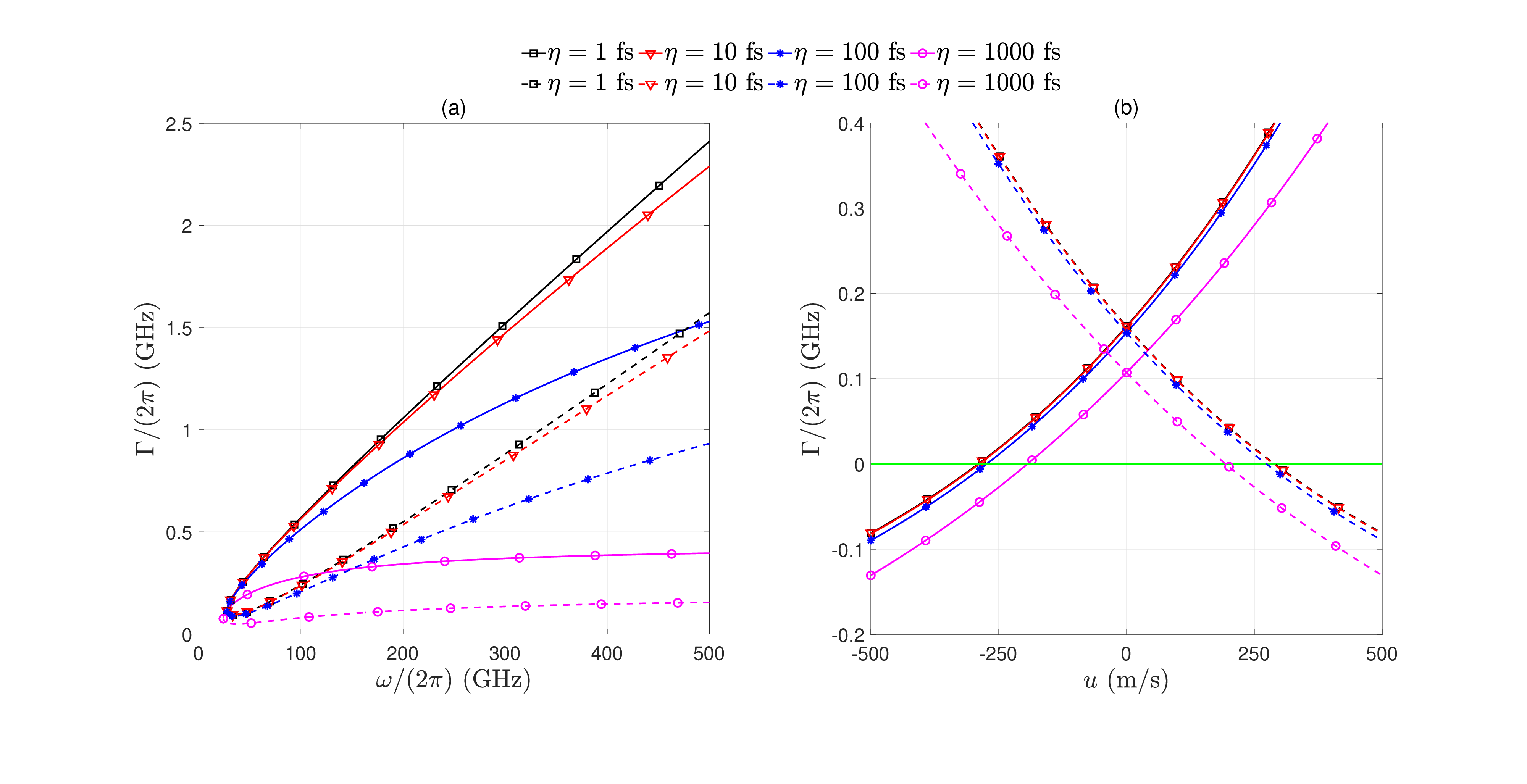}
\caption{(color online). Dependence of the damping factor $\Gamma$ on $\omega$ (a) and $u$ (b) for the precessional spin waves. In (a), the solid and dashed lines correspond to $u = 100 $m$/$s and $- 100 $m$/$s, respectively. In (b), $\omega/(2 \pi) = 40$ GHz. the solid and dashed lines correspond to $k > 0$ and $k < 0$, respectively. Other parameters are the same as those in Fig. \ref{omega_and_Gamma_vs_k}.} \label{Gamma_vs_omega_and_u}
\end{figure*}

Now, let us continue to discuss the dissipation of spin waves based on Eqs. (\ref{re_eq_tem}, \ref{im_eq_tem}). The damping factor is related to the linewidth ($\vert\Gamma\vert/\pi$) in resonance experiments \cite{RozsaL,LomonosovAM}. Since magnetic inertia influences $\Gamma$, it is possible to indirectly measure $\eta$ through spin-wave resonance experiments that utilize nonreciprocity. Given that the precessional spin waves are routinely excited and probed, we pay attention to the dependence of $\Gamma$ on both $\omega$ and $u$. Our idea is illustrated in Fig.~\ref{Gamma_vs_omega_and_u}(a), which shows the variation of $\Gamma$ with $\omega$ for both positive and negative current directions $u$. It is intriguing that the effective damping of spin waves $\Gamma$ for $u > 0$ is obviously larger than that for $u < 0$. This feature can be exploited to determine the inertial relaxation time $\eta$ of ferromagnets. By reversing the current, one can measure the difference $\Gamma(u) - \Gamma(-u)$ for varying $\omega$. Then, $\eta$ may be determined by matching the experimental result with the theoretic curve for a certain $\eta$. It is worth mentioning that the subtraction of $\Gamma$ for opposite current may remove other contributions, such as inhomogeneous broadening.

Let us proceed again to the specific case of $N_y = N_z$, for which, it is feasible to analytically calculate the difference in $\Gamma$ for two opposite current directions. By substituting $\omega^+_p$ from Eq. (\ref{omega_p}) into $\Gamma^+_p$ [Eq. (\ref{Gamma_p})] and applying the approximations $1 + \alpha^{2} \approx 1$ and $1 + \alpha \beta \approx 1$, we derive the expression for $\Delta \Gamma = \Gamma(u) - \Gamma(- u)$ for the precessional spin wave propagating in the positive direction ($k > 0$),
\begin{equation}
\Delta \Gamma = \frac{\beta u \sqrt{4 \left( \gamma J \chi^2 + \eta \beta^2 u^2 \right) \left( \omega - \omega_0 + \eta \omega^2 \right) + \chi^2 u^2}}{\gamma J \chi^2 + \eta \beta^2 u^2}, \label{DelGam}
\end{equation}
where $\chi = 1 + 2 \eta \omega$, and $\omega_0 = \gamma (1/2 M_s + H_k) - u^2/(4 \gamma J)$. Eq. (\ref{DelGam}) indicates that the difference $\Delta \Gamma$ becomes larger at higher frequencies. By utilizing the parameters as Fig. \ref{omega_and_Gamma_vs_k}, it is easy to infer that $\gamma J \chi^2 \gg \eta \beta^2 u^2$. Hence, $\Delta \Gamma$ is almost proportional to $\beta$. This implies that the nonadiabaticity of STTs enhances the ability to accurately distinguish $\Gamma(u)$ and $\Gamma(- u)$.

Besides the frequency $\omega$, the STT parameter $u$, which scales linearly with the current density, is also adjustable. So, we display the dependence of $\Gamma$ on $u$ in  Fig. \ref{Gamma_vs_omega_and_u}(b). When the current direction aligns with the spin-wave propagation direction, $\Gamma$ increases with $u$. Conversely, when the current opposes the spin-wave propagation, $\Gamma$ decreases with $u$ and can even change sign at a critical value $u_{c}$. Note that the negative damping factor means the amplification of spin waves by the STTs.

To find the critical current $u_{c}$, at which the damping factor vanishes $\Gamma = 0$, we apply some reasonable approximations. First we assume that $\Gamma^2 \ll \omega^2$ based on the comparison of numerical data shown in Fig. \ref{omega_and_Gamma_vs_k}(a) and Fig. \ref{omega_and_Gamma_vs_k}(b). This approximation is also justified by the smallness of damping which leads to a much longer characteristic damping time compared to the oscillation period. Then solving Eq. (\ref{im_eq_tem}) yields
\begin{equation}
\Gamma = \frac{b - \sqrt{b^2 - 4 a c}}{2 a}, \label{Gamma}
\end{equation}
where
\begin{eqnarray}
a &=& 2 \alpha \eta \omega, \\
b &=& 2 \eta \omega \left( \frac{\omega_y + \omega_z}{2} - \eta \omega^2 \right) + \omega + u k, \label{par_b2} \\
c &=& \left( \frac{\omega_y + \omega_z}{2} - \eta \omega^2 \right) \left( \alpha \omega + \beta u k \right). \label{par_c}
\end{eqnarray}
In Eq. (\ref{par_b2}), we employ next approximations that $1 + \alpha \beta \approx 1$ and $1 + \alpha^2 \approx 1$. From Eq. (\ref{Gamma}), it can be inferred that $c = 0$ defines a transition point, across which $\Gamma$ changes sign. At this point, $\alpha \omega + \beta u_c k = 0$ and $\Gamma = 0$. Then, combining with Eq. (\ref{re_eq_tem}), the critical $u$ is derived as
\begin{equation}
u_c = - \frac{\alpha}{\beta} \omega \sqrt{\frac{\gamma J}{\eta \omega^2 + \sqrt{\left( \frac{\omega^0_y - \omega^0_z}{2} \right)^2 + \left( 1 - \frac{\alpha}{\beta} \right)^2 \omega^2} - \frac{\omega^0_y + \omega^0_z}{2}}}, \label{cri_u}
\end{equation}
where $\omega^0_{y,z} = \gamma (H_k + N_{y,z} M_s)$. The competition between the nonadiabatic STT $\beta$ and the intrinsic damping $\alpha$ determines both the critical current $u_c$ and the effective damping $\Gamma$. Furthermore, $\eta$ appears only in the term $\eta \omega^2$ in the square root of Eq. (\ref{cri_u}). Thus, at higher frequencies, magnetic inertia has a greater influence on $u_c$, which decreases as $\eta$ increases.

\section{spatial attenuating spin waves} \label{spa}

\begin{figure*}[th]
\includegraphics[scale=0.26,angle=0,trim=0cm 3cm 0cm 0cm]{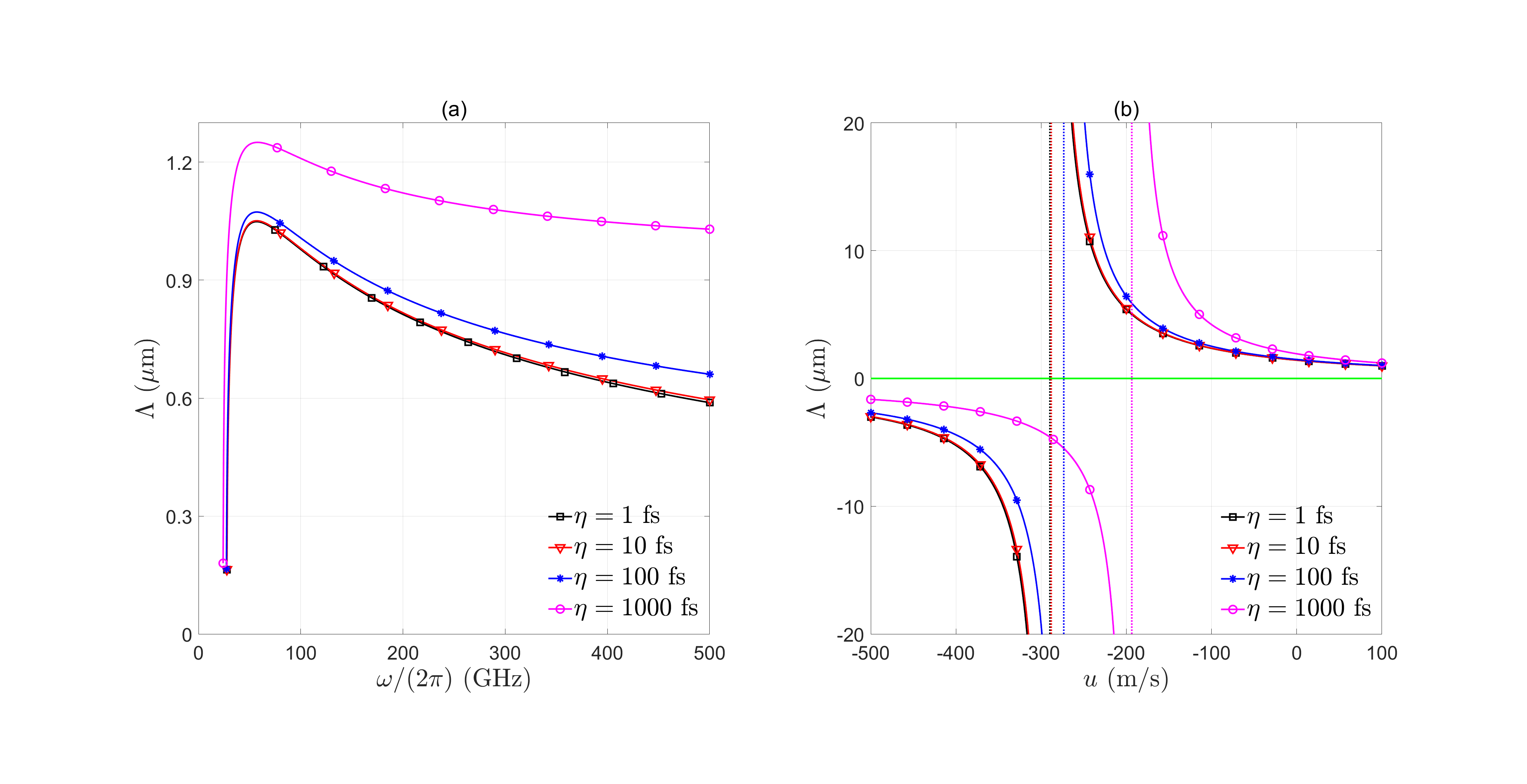}
\caption{(color online). Dependence of the attenuating length $\Lambda$ on $\omega$ (a) and $u$ (b) for the precessional spin waves. For (a), $u = 100 $m$/$s. For (b), $\omega/(2 \pi) = 40$ GHz. Other parameters are the same as those in Fig. \ref{omega_and_Gamma_vs_k}.} \label{Lambda_vs_omega_and_u}
\end{figure*}

In spin-wave transport experiments, the focus is on the spatial attenuation of spin waves. Therefore, in the spin-wave ansatz (\ref{ansatz}), it is reasonable to assume that $\tilde{\omega} = \omega$ is real, representing the frequency, and
\begin{equation}
\tilde{k} = k + i \frac{1}{\Lambda}, \label{complex_k}
\end{equation}
where $k$ is the magnitude of the wave vector and $\Lambda$ is the attenuating length, which can be measured experimentally \cite{ChauleauJY,WangJ}. Following the same procedure as for deriving Eqs. (\ref{re_eq_tem}, \ref{im_eq_tem}), we obtain two higher-order polynomial equations in terms of $\omega$ and $\Lambda$,
\begin{small}
\begin{eqnarray}
&& \left[ \frac{\omega_y + \omega_z}{2} - \frac{\gamma J}{\Lambda^2} + \frac{\beta u}{\Lambda} - \eta \omega^2 \right]^2 - \left[ \left( \alpha \omega + \beta u k \right) - 2 \frac{\gamma J k}{\Lambda} \right]^2 \notag \\ && + \frac{u^2}{\Lambda^2} - \left( \omega + u k \right)^2 = \left[ \frac{\omega_y - \omega_z}{2} \right]^2, \label{re_eq_spa} \\
&& \left[ \frac{\omega_y + \omega_z}{2} - \frac{\gamma J}{\Lambda^2} + \frac{\beta u}{\Lambda} - \eta \omega^2 \right]  \left[ \left( \alpha \omega + \beta u k \right) - 2 \frac{\gamma J k}{\Lambda} \right] \notag \\ && + \frac{u}{\Lambda} \left( \omega + u k \right) = 0, \label{im_eq_spa}
\end{eqnarray}
\end{small}
where $\omega_{y,z}$ are defined in Eq. (\ref{omega_yz_2}). The structure of Eqs. (\ref{re_eq_spa}, \ref{im_eq_spa}) is similar to that of Eqs. (\ref{re_eq_tem}, \ref{im_eq_tem}). Since, there is no close-form solution for the general case when $\omega_y \neq \omega_z$, we solve them numerically and plot the variations of $\Lambda$ with $\omega$ and $u$ for precessional spin waves in Fig. \ref{Lambda_vs_omega_and_u}. For clarity, we restrict our discussion to the case of $\omega > 0$ and $k >0$, meaning spin waves propagating along the positive $x$ direction. In Fig. \ref{Lambda_vs_omega_and_u}(a), a moderate value of $u$ is selected, and $\omega$ starts from the resonance frequency $\omega_p$ [see Eq. (\ref{eigen_fre})], below which the spin waves can not propagate \cite{SeoSM}. As $\omega$ increases, the attenuating length grows rapidly, reaches a maximum, and then decreases gradually. Moreover, the stronger magnetic inertia, the longer attenuating length.

The numerically calculated Fig. \ref{Lambda_vs_omega_and_u}(b) illustrates the dependance of $\Lambda$ on $u$ for a fixed frequency and different inertial relaxation times. When the current is opposite to the spin wave ($u < 0$), $\Lambda$ flips sign at a critical value $u_c$ as $\vert u \vert$ increases. This indicates that the spin wave is amplified when $u < u_c$, as the nonadiabatic STT overcomes the intrinsic damping. Additionally, the attenuating length approaches infinity as $u$ nears $u_c$. By tuning frequency $\omega$ or $u$, and measuring $\Lambda$ \cite{WangJ}, one can determine the inertial relaxation time $\eta$ by leveraging the $\eta$-dependance of the theoretical $\Lambda$-$\omega$ or $\Lambda$-$u$ curves. This approach offers another experimental pathway for investigating magnetic inertia.

To shed light on the physical origin of the transition of the attenuating length $\Lambda$ at critical $u_c$ in inertial ferromagnets, we assume small Gilbert damping and nonadiabaticity, $\alpha \beta \ll 1$ and $\beta^2 \ll 1$. Additionally, we adopt the assumption that the attenuation length is much larger than the wavelength, i.e. $1/(k^2 \Lambda^2) \ll 1$. Under these approximations, $\Lambda$ can be derived from Eq. (\ref{im_eq_spa}) as:
\begin{equation}
\Lambda = \frac{2 a^\prime}{\sqrt{b^{\prime 2} + 4 a^\prime c} - b^\prime}, \label{att_len}
\end{equation}
where
\begin{eqnarray}
a^\prime &=& 2 \gamma J \beta u k, \\
b^\prime &=& 2 \gamma J k \left[ \frac{\omega_y + \omega_z}{2} - \eta \omega^2 \right] - u \left( \omega + u k \right), \label{par_b}
\end{eqnarray}
and $c$ is given by Eq. (\ref{par_c}). From Eq. (\ref{att_len}), it can be inferred that $c = 0$ defines a transition point at which $\Lambda$ changes sign. At this point $\alpha \omega + \beta u_c k = 0$ and $1/\Lambda = 0$, indicating a critical current where the attenuation length becomes infinite and the spin wave experiences no damping. The term $\alpha \omega$ arises from the damping term in ILLG equation (\ref{iLLG}), while $\beta u_c k$ originates from the nonadiabatic STT, which acts as an anti-damping force, depending on the current direction. When a balance is achieved between the damping and anti-damping effects, net energy dissipation vanishes, and the spin-wave attenuation is suppressed.

It is possible to derive the critical value $u_c$ from Eqs. (\ref{re_eq_spa}, \ref{att_len}), which turns out to be the same as for temporally damped spin waves [see Eq. (\ref{cri_u})], as expected. This consistency confirms that the critical current value $u_c$ governs both the temporal and spatial attenuation of spin waves. Note that in the absence of magnetic inertia ($\eta = 0$), the expression for $u_c$ aligns with the one derived by Seo et al. \cite{SeoSM}.

\section{Conclusion} \label{con}

In this paper, we investigate the dispersion and dissipation of inertial spin waves in ferromagnetic strips driven by STTs. For different scenarios, two types of spin-wave ansatz are employed, describing the exponential time and
space decays, characterized by the damping factor $\Gamma$ and the attenuating length $\Lambda$, respectively. Both parameters are experimentally measurable. We find that the adiabatic STT shifts the dispersion curves along the frequency and wave vector axes, displaying typical features of the Doppler effect. The nonadiabatic STT, acting as a damping or anti-damping mechanism, can either weaken or enhance spin-wave decay, depending on the current
direction. Notably, we show that magnetic inertia significantly modifies the dependence of $\Gamma$ and $\Lambda$ on both the frequency of the wave source and the current driving the STTs. These findings suggest that the inertial relaxation time can be experimentally determined by tuning the frequency or current and measuring $\Gamma$ and $\Lambda$.

\section{Acknowledgments}

Peng-Bin He is supported by the NSF of Changsha City (Grant No. kq2208008) and the NSF of Hunan Province (Grant No. 2023JJ30116). We are grateful to Igor Barsukov for fruitful discussions.


\begin{thebibliography}{99}

\bibitem{NeerajK} K. Neeraj, N. Awari, S. Kovalev, D. Polley, N. Z. Hagstr\"{o}m, S. S. P. K. Arekapudi, A. Semisalova, K. Lenz, B. Green, J.-C. Deinert, I. Ilyakov, M. Chen, M. Bawatna, V. Scalera, M. d'Aquino, C. Serpico, O.
    Hellwig, J.-E. Wegrowe, M. Gensch, and S. Bonetti, Inertial spin dynamics in ferromagnets, Nat. Phys. \textbf{17}, 245 (2021).
\bibitem{LiY} Y. Li, A.-L. Barra, S. Auffret, U. Ebels, and W. E. Bailey, Inertial terms to magnetization dynamics in ferromagnetic thin films, Phys. Rev. B \textbf{92}, 140413(R) (2015).
\bibitem{UnikandanunniV} V. Unikandanunni, R. Medapalli, M. Asa, E. Albisetti, D. Petti, R. Bertacco, E. E. Fullerton, and S. Bonetti, Inertial spin dynamics in epitaxial cobalt films, Phys. Rev. Lett. \textbf{129}, 237201 (2022).
\bibitem{DeA} A. De, J. Schlegel, A. Lentfert, L. Scheuer, B. Stadtm\"{u}ller, P. Pirro, G. von Freymann, U. Nowak, and M. Aeschlimann, Nutation: separating the spin from its magnetic moment, arXiv, 2405.01334 (2024).

\bibitem{MondalR_JMMM} R. Mondal, L. R\'{o}zsa, M. Farle, P. M. Oppeneer, U. Nowak, and M. Cherkasskii, Inertial effects in ultrafast spin dynamics, J. Magn. Magn. Mater. \textbf{579}, 170830 (2023).
\bibitem{OliveE_APL} E. Olive, Y. Lansac, and J.-E. Wegrowe, Beyond ferromagnetic resonance: The inertial regime of the magnetization, Appl. Phys. Lett. \textbf{100}, 192407 (2012). 
\bibitem{BottcherD} D. B\"{o}ttcher and J. Henk, Significance of nutation in magnetization dynamics of nanostructures, Phys. Rev. B \textbf{86}, 020404 (2012).
\bibitem{OliveE_JAP} E. Olive, Y. Lansac, M. Meyer, M. Hayoun, and J.-E. Wegrowe, Deviation from the Landau-Lifshitz-Gilbert equation in the inertial regime of the magnetization, J. Appl. Phys. \textbf{117}, 213904 (2015). 
\bibitem{CherkasskiiM_PRB102} M. Cherkasskii, M. Farle, and A. Semisalova, Nutation resonance in ferromagnets, Phys. Rev. B \textbf{102}, 184432 (2020). 
\bibitem{TitovSV_PRB103} S. V. Titov, W. T. Coffey, Y. P. Kalmykov, and M. Zarifakis, Deterministic inertial dynamics of the magnetization of nanoscale ferromagnets, Phys. Rev. B \textbf{103}, 214444 (2021).
\bibitem{CiorneiMC} M.-C. Ciornei, J. M. Rub\'{i}, and J.-E. Wegrowe, Magnetization dynamics in the inertial regime: Nutation predicted at short time scales, Phys. Rev. B \textbf{83}, 020410(R) (2011). 
\bibitem{WegroweJE} J.-E. Wegrowe and M.-C. Ciornei, Magnetization dynamics, gyromagnetic relation, and inertial effects, Am. J. Phys. \textbf{80}, 607 (2012). 

\bibitem{WinterL} L. Winter, S. Gro{\ss}enbach, U. Nowak , and L. R\'{o}zsa, Nutational switching in ferromagnets and antiferromagnets, Phys. Rev. B \textbf{106}, 214403 (2022).

\bibitem{CherkasskiiM_PRB106} M. Cherkasskii, I. Barsukov, R. Mondal, M. Farle, and A. Semisalova, Theory of inertial spin dynamics in anisotropic ferromagnets, Phys. Rev. B \textbf{106}, 054428 (2022). 
\bibitem{TitovSV_JAP} S. V. Titov, W. J. Dowling, and Y. P. Kalmykov, Ferromagnetic and nutation resonance frequencies of nanomagnets with various magnetocrystalline anisotropies, J. Appl. Phys. \textbf{131}, 193901 (2022). 

\bibitem{KimelAV} A. V. Kimel, B. A. Ivanov, R. V. Pisarev, P. A. Usachev, A. Kirilyuk, and T. Rasing, Inertia-driven spin switching in antiferromagnets, Nat. Phys. \textbf{5}, 727 (2009).
\bibitem{RodriguezR} R. Rodriguez, M. Cherkasskii, R. Jiang, R. Mondal, A. Etesamirad, A. Tossounian, B. A. Ivanov, and I. Barsukov, Spin Inertia and Auto-Oscillations in Ferromagnets, Phys. Rev. Lett. \textbf{132}, 246701 (2024).
\bibitem{MondalR_PRB103} R. Mondal, S. Gro{\ss}enbach, L. R\'{o}zsa, and U. Nowak, Nutation in antiferromagnetic resonance, Phys. Rev. B \textbf{103}, 104404 (2021). 
\bibitem{MondalR_PRB104} R. Mondal and P. M. Oppeneer, Influence of intersublattice coupling on the terahertz nutation spin dynamics in antiferromagnets, Phys. Rev. B \textbf{104}, 104405 (2021). 
\bibitem{CherkasskiiM_PRB109} M. Cherkasskii, R. Mondal, and L. R\'{o}zsa, Inertial spin waves in spin spirals, Phys. Rev. B \textbf{109}, 184424 (2024).

\bibitem{BhattacharjeeS} S. Bhattacharjee, L. Nordstr\"{o}m, and J. Fransson, Atomistic Spin Dynamic Method with both Damping and Moment of Inertia Effects Included from First Principles, Phys. Rev. Lett. \textbf{108}, 057204 (2012).
\bibitem{ThonigD} D. Thonig, O. Eriksson, and M. Pereiro, Magnetic moment of inertia within the torque-torque correlation model, Sci. Rep. \textbf{7}, 931 (2017).

\bibitem{FlebusB} B. Flebus, D. Grundler, B. Rana, Y. Otani, I. Barsukov, A. Barman, G. Gubbiotti, P. Landeros, J. Akerman, U. S. Ebels, and others, The 2024 magnonics roadmap, J. Phys.: Condens. Matter. \textbf{36}, 363501 (2024).

\bibitem{SeoSM} S.-M. Seo, K.-J. Lee, H. Yang, and T. Ono, Current-Induced Control of Spin-Wave Attenuation, Phys. Rev. Lett. \textbf{102}, 147202 (2009).
\bibitem{MoonJH} J.-H. Moon, S.-M. Seo, K.-J. Lee, K.-W. Kim, J. Ryu, H.-W. Lee, R. D. McMichael, and M. D. Stiles, Spin-wave propagation in the presence of interfacial Dzyaloshinskii-Moriya interaction, Phys. Rev. B \textbf{88}, 184404 (2013). 
\bibitem{KimDH} D.-H. Kim, S.-H. Oh, D.-K. Lee, S. K. Kim, and K.-J. Lee, Current-induced spin-wave Doppler shift and attenuation in compensated ferrimagnets, Phys. Rev. B \textbf{103}, 014433 (2021). 

\bibitem{ZhouZW} Z.-W. Zhou, X.-G. Wang, Y.-Z. Nie, Q.-L. Xia, Z.-M. Zeng, and G.-H. Guo, Left-handed polarized spin waves in ferromagnets induced by spin-transfer torque, Phys. Rev. B \textbf{99}, 014420 (2019).

\bibitem{ChauleauJY} J.-Y. Chauleau, H. G. Bauer, H. S. K\"{o}rner, J. Stigloher, M. H\"{a}rtinger, G. Woltersdorf, and C. H. Back, Self-consistent determination of the key spin-transfer torque parameters from spin-wave Doppler experiments, Phys. Rev. B \textbf{89}, 020403(R) (2014). 

\bibitem{XingXJ} X. J. Xing, Y. P. Yu, and S. W. Li, Modulation of propagation characteristics of spin waves induced by perpendicular electric currents, Appl. Phys. Lett. \textbf{95}, 142508 (2009).
\bibitem{WooS} S. Woo and G. S. D. Beach, Control of propagating spin-wave attenuation by the spin-Hall effect, J. Appl. Phys. \textbf{122}, 093901 (2017).

\bibitem{YanZM} Z.-M. Yan, Z.-X. Li, X.-G. Wang, Z.-Y. Luo, Q.-L Xia, Y.-Z Nie, and G.-H Guo, Manipulation of spin-wave attenuation and polarization in antiferromagnets, Phys. Rev. B \textbf{108}, 134432 (2023). 

\bibitem{ManagoT} T. Manago, K. Yamanoi, S. Kasai, and S. Mitani, Damping factor estimation using spin wave attenuation in permalloy film, J. Appl. Phys. \textbf{117}, 17D121 (2015).

\bibitem{MakhfudzI_APL} I. Makhfudz, E. Olive, and S. Nicolis, Nutation wave as a platform for ultrafast spin dynamics in ferromagnets, Appl. Phys. Lett. \textbf{117}, 132403 (2020). 
\bibitem{CherkasskiiM_PRB103} M. Cherkasskii, M. Farle, and A. Semisalova, Dispersion relation of nutation surface spin waves in ferromagnets, Phys. Rev. B \textbf{103}, 174435 (2021). 
\bibitem{LomonosovAM} A. M. Lomonosov, V. V. Temnov, and J.-E. Wegrowe, Nutation spin waves in ferromagnets, Phys. Rev. B \textbf{104}, 054425 (2021). 
\bibitem{TitovSV_PRB105} S. V. Titov, W. J. Dowling, Y. P. Kalmykov, and M. Cherkasskii, Nutation spin waves in ferromagnets, Phys. Rev. B \textbf{105}, 214414 (2022). 
\bibitem{MondalR_PRB106} R. Mondal, and L. R\'{o}zsa, Inertial spin waves in ferromagnets and antiferromagnets, Phys. Rev. B \textbf{106}, 134422 (2022). 

\bibitem{HePB_PRB108} P.-B. He, Large-amplitude and widely tunable self-oscillations enabled by the inertial effect in uniaxial antiferromagnets driven by spin-orbit torques, Phys. Rev. B \textbf{108}, 184418 (2023).
\bibitem{HePB_PRB110} P.-B. He, Influence of the magnetic inertia on the self-oscillation in spin-orbit torque-driven tripartite antiferromagnets with a $120^\circ$ rotation symmetry, Phys. Rev. B \textbf{110}, 064411 (2024).

\bibitem{WangJ} J. Wang, H. Wang, J. Chen, W. Legrand, P. Chen, L. Sheng, J. Xia, G. Lan, Y. Zhang, R. Yuan, J. Dong, X. Han, J.-P. Ansermet, and H. Yu, Broad-wave-vector spin pumping of flat-band magnons, Phys. Rev. Appl. \textbf{21}, 044024 (2024). 

\bibitem{ZhangS} S. Zhang and Z. Li, Roles of Nonequilibrium Conduction Electrons on the Magnetization Dynamics of Ferromagnets, Phys. Rev. Lett. \textbf{93}, 127204 (2004).

\bibitem{AharoniA} A. Aharoni, Demagnetizing factors for rectangular ferromagnetic prisms, J. Appl. Phys. \textbf{83}, 3432 (1998).

\bibitem{VlaminckV} V. Vlaminck and M. Bailleul, Current-Induced Spin-Wave Doppler Shift, Science, \textbf{322}, 410 (2008).
\bibitem{SugimotoS} S. Sugimoto, M. C. Rosamond, E. H. Linfield, and C. H. Marrows, Observation of spin-wave Doppler shift in Co$_{90}$Fe$_{10}/$Ru micro-strips for evaluating spin polarization, Appl. Phys. Lett. \textbf{109}, 112405 (2016). 

\bibitem{LiuX} X. Liu, R. Sooryakumar, C. J. Gutierrez, and G. A. Prinz, Exchange stiffness and magnetic anisotropies in bcc Fe$_{1-x}$Co$_x$ alloys, J. Appl. Phys. \textbf{75}, 7021 (1994).
\bibitem{WeberR} R. Weber, D.-S. Han, I. Boventer, S. Jaiswal, R. Lebrun, G. Jakob, and M. Kl\"{a}ui, Gilbert damping of CoFe-alloys, J. Phys. D: Appl. Phys. \textbf{52} 325001 (2019).

\bibitem{RozsaL} L. R\'{o}zsa, J. Hagemeister, E. Y. Vedmedenko, and R. Wiesendanger, Effective damping enhancement in noncollinear spin structures, Phys. Rev. B \textbf{98}, 100404 (2018).

\end{thebibliography}
\end{document}